\documentclass[final]{beatcs}
\usepackage{amssymb, amsfonts}
\usepackage[english]{babel}
\usepackage{verbatim}

\newtheorem{theorem}{Theorem}

\newtheorem{lemma}{Lemma}

\newcommand{\NI}{\noindent}
\newcommand{\HB}{\hfill{$\Box$}}
\newcommand{\III}{\vspace{3 mm}}

\title{The H-index can be easily manipulated}

\author{Bart de Keijzer \thanks{Centre for Mathematics and Computer Science (CWI), \url{keijzer@cwi.nl}}
\and
Krzysztof R. Apt \thanks{Centre for Mathematics and Computer Science (CWI) and ILLC, University of Amsterdam, The Netherlands, \url{apt@cwi.nl}}
}

\begin{document}
\pagestyle{plain}
\bibliographystyle{abbrv}
\sloppy 

\maketitle

\begin{abstract}
  We prove two complexity results about the H-index concerned with the
  Google scholar \emph{merge} operation on one's scientific articles. The results show that, although it is hard to merge one's articles in an optimal way, it is easy to merge them in such a way that one's H-index increases. This suggests the need for an alternative scientific performance measure that is resistant to this type of manipulation.
\end{abstract}

\section{Introduction}

The \emph{H-index} was introduced by the physicist J.E. Hirsch
in~\cite{Hir05} to `quantify an individual's scientific research
output'. Recall that it is defined as the largest $x$ such that one's
$x$ most cited paper is cited at least $x$ times. (An aside: Hirsch's
original definition was ambiguous as pointed out in~\cite{Ste07}, where the
current definition is proposed.) Its introduction
led to an impressive literature. According to Google scholar; by 18th
of April 2013 this paper was cited 3043 times. To mention just
one example, \cite{Woe08} provided its axiomatic definition.

The H-index started to be used as a universal measure to assess
and compare researchers in a given discipline.  Hirsch suggested in
his paper `(with large error bars) that for faculty at major research
universities, $h \approx 12$ might be a typical value for advancement
to tenure (associate professor) and that $h \approx 18$ might be a
typical value for advancement to full professor'.

In fact, computer scientists seem to cite each other much more
often. Jens Palsberg maintains at
\url{http://www.cs.ucla.edu/~palsberg/h-number.html} a list of
computer scientists with H-index 40 or higher (a value corresponding in
Hirsch's article to Nobel prize winners). The list has more than
600 names and is based on the output generated by Google scholar.

Several people made obvious observations that the H-index can be
boosted by such simple measures as adding your name to the articles
written by members of your group, splitting a long article into a
couple of shorter ones, by citing one's and each other's work, etc.
For example, \cite{BK11a} studies the problem of manipulability of the
H-index by means of self-citations.  

This brings us to the subject of this note. \emph{Google scholar}
allows one to perform some operations on the listed articles; notably, the
\emph{merge}-operation allows one to combine two versions of an article even
if they have different titles.  By means of the merge operation, you
can obviously improve your H-index. Suppose for instance that your
H-index is 20. Then you can increase it by merging two articles that are
cited each 11 times.

This suggests two natural problems, where in each case we refer to
the improvement of the H-index by means of the merge operation.

\begin{itemize}
\item Is it possible to improve your H-index?

\item Given a number $k$, determine whether
your H-index can be improved to at least $k$.

\end{itemize}

\section{Two results}

To deal with these questions, we introduce first some notation. A researcher's output is
represented as a multiset of natural numbers, each number representing a publication and its value
representing the number of its citations. For example the multiset $\{1,1,2,3,4,4,5,5,5\}$
represents an output consisting of 9 publications with the corresponding H-index 4.
Given a multiset $T$ of numbers we abbreviate $\sum_{x \in T} x$ to $\sum T$. 
So $\sum T$ is the number of citations resulting from the merge of the publications in $T$ into one.

To deal with the outcomes of merges we need to consider partitions of such multisets.

Fix a finite multiset $S$ of numbers from $\mathbb{N}_{> 0}$.
We denote by $\bar{S}$ the singletons partition $\{\{x\}\ |\ x \in S\}$. 
Given a partition $\mathcal{T}$ of $S$, we define 
\begin{equation*}
v(\mathcal{T}) = \max\{|\mathcal{T}'|\ |\ \mathcal{T}' \subseteq \mathcal{T}, \forall T \in \mathcal{T'} : \sum T \geq |\mathcal{T}'| \},
\end{equation*}
where, as usual, $|\mathcal{T}'|$ denotes the cardinality of the multiset $\mathcal{T}'$ (which is a submultiset of a partition of $S$ in this case).
In words, call a subset $\mathcal{T}'$ of the partition $\mathcal{T}$
\emph{good} if each element $T$ of $\mathcal{T}'$ after merge into a
single publication yields at least $|\mathcal{T}'|$ citations. So if
one allows the merge operation, then a good partition $\mathcal{T}'$
ensures that the H-index can be set to at least $|\mathcal{T}'|$.
Then $v(\mathcal{T})$ is the cardinality of the largest good subset of
$\mathcal{T}$, hence $v(\mathcal{T})$ is the largest H-index one can
obtain by means of the merge operation, while $v(\bar{S})$ is the
H-index corresponding to the input multiset $S$. To put it more
directly,
\[
v(\bar{S}) = \max\{|T|\ |\ T \subseteq S, \ \forall x \in T \: x \geq |T| \},
\]
where we refer to the submultisets.

We call a partition $\mathcal{S}$ of $S$ an \emph{improving partition}
if $v(\mathcal{S}) > v(\bar{S})$.
We can now formalize the above two problems as follows, given as input 
a finite multiset $S$ of numbers in $\mathbb{N}_{> 0}$. 

\paragraph{H-index improvement problem} Does there 
exist an improving partition? If yes, find it.

\paragraph{H-index achievability problem} Given a number $k$, does there exist a partition $\mathcal{T}$ of $S$, such that $v(\mathcal{T}) \geq k$?
~\\
\\
In Section \ref{sect:proofs}, we present the proofs of the following two results.

\begin{theorem}
\label{thm:one}
The H-index improvement problem can be solved in polynomial time.
\end{theorem}

\begin{theorem}
\label{thm:two}
The H-index achievability problem is strongly $\mathsf{NP}$-complete.\footnote{A decision problem that involves numerical input is said to be \emph{strongly} $\mathsf{NP}$-complete if the problem is $\mathsf{NP}$ complete even if all the numbers in the input are represented in unary.}
\end{theorem}

In particular, it is strongly $\mathsf{NP}$-hard to compute the maximal H-index that can be achieved
through the merge operation.

From the viewpoint of manipulability, Theorem~\ref{thm:one} is bad
news. Ideally, we would like to have a performance measure that is
computationally difficult to manipulate. One can see a parallel with the
search for voting methods that are difficult to manipulate, see, e.g.
\cite{ZFBE12}.
Our conclusion is that the
H-index is not the last word in the ongoing quest to find a credible
way to quantify one's scientific output.

\section{Proofs of the theorems}\label{sect:proofs}

In what follows, we assume that a multiset is represented as a list of
possibly duplicate numbers. A different way of representing a multiset
would be the more compact one, where we list only the distinct numbers
that appear in the multiset, along with their respective multiplicity.
We consider the latter representation to be unnatural, given the
context in which we study this problem.  
\III

\NI
\textbf{Proof of Theorem~\ref{thm:one}}.
Let $S$ be the given multiset.  Let $S'$ be the smallest submultiset
of $S$ such that $v(\bar{S}) = v(\overline{S'})$. For instance, if $S =
\{5, 4, 3, 3, 3, 2\}$, then $S' = \{5, 4, 3\}$ and if
$S = \{5, 3, 3, 3, 3, 2\}$, then $S' = \{5, 3, 3\}$. In both cases $v(\bar{S}) = 3$.
Call a number $x \in S'$ \emph{supercritical} if $x > v(\bar{S})$ and
\emph{critical} if $x = v(\bar{S})$.  Let $C_{+}$ be the multiset of
all supercritical numbers in $S'$ and $C$ the multiset of all critical
numbers in $S'$. Note that $C$ and $C_{+}$ partition $S'$ and that $v(\bar{S}) = |C_{+}| + |C|$.
Furthermore, let $L$ denote the multiset of $|C|$ smallest numbers in
$S$.

For instance, if $S = \{5, 4, 3, 3, 3, 2\}$, then $C = \{3\}$ and $L = \{2\}$,
and if $S = \{5, 3, 3, 3, 3, 2\}$, then $C = \{3, 3\}$ and $L = \{3, 2\}$.

Note that below, we treat duplicate numbers in $S$ as having ``separate identities'', 
so that for two numbers $x,y \in S$ that are equal in magnitude, it may hold that $x \in C$ but $y \not\in C$ or $x \in L$ but $y \not\in L$.
We believe that this slight informality and definitional abuse will cause no confusion to the reader.

We first establish the following characterization result.

\begin{lemma}
 There exists an improving partition of $S$ iff $L \cap C = \varnothing$ and $\sum S \backslash (C
\cup C_{+} \cup L) > |C| + |C_{+}|$.
\end{lemma}

\NI
\emph{Proof}. Suppose there exists an improving partition $\mathcal{S}$ of $S$.

We can assume without loss of generality that 
the following properties then hold:

\begin{enumerate}
  
\item Each supercritical number in $S$ appears in a singleton set in
  $\mathcal{S}$. These are the only singleton sets in $\mathcal{S}$.
  
  Indeed, if a supercritical number $x \in S$ appears in a
  non-singleton set $T \in \mathcal{S}$, then take the partition
  $\mathcal{T}$ of $S$ obtained from $\mathcal{S}$ by splitting $T$
  into singletons. Because $\mathcal{S}$ is an improving partition,
  there are at least $v(\bar{S})$ multisets $T' \in \mathcal{S}
  \backslash \{T\}$ such that $\sum T' > v(\bar{S})$. All multisets of
  $\mathcal{S} \backslash \{T\}$ are in $\mathcal{T}$. Also the number
  $x$ is in a singleton set of $\mathcal{T}$ and $x > v(\bar{S})$.
  Therefore, there are in $\mathcal{T}$ at least $v(\bar{S}) + 1$
  multisets $T'$ such that $\sum T' > v(\bar{S})$. Hence,
  $\mathcal{T}$ is an improving partition.
  
  After we have repeatedly performed the above splitting steps we
  obtain an improving partition $\mathcal{S'}$ such that each
  supercritical number $x \in S$ appears in a singleton set in
  $\mathcal{S'}$.

Since
\[
v(\mathcal{S'}) > v(\bar{S}) = |C_{+}| + |C| \geq |C_{+}|,
\]
there exists in $\mathcal{S'}$ a non-singleton multiset $T \in
\mathcal{S}$ that contains only non-supercritical numbers. Merging
with it all singleton sets that contain a non-supercritical number
yields the desired improving partition.

\item $L$ is disjoint from $C$. 
  
  By Property 1, the supercritical numbers form singleton sets in
  $\mathcal{S}$, and each remaining multiset has cardinality at least
  $2$. If $L$ were not disjoint from $C$, then we would have $|S| \leq
  |C_{+}| + |L| + |C|$, so $|S \backslash C_{+}| \leq |L| + |C| =
  2|C|$, hence the number $\ell$ of non-singleton multisets in
  $\mathcal{S}$ would be at most $|C|$. This yields a contradiction,
  since we would then have $v(\mathcal{S}) \leq |C_{+}| + \ell \leq
  |C_{+}|+|C| = v(\bar{S})$.

\item In $\mathcal{S}$, every critical number is in a set of cardinality 2.
  
  Indeed, by Property 1, critical numbers do not appear in singleton
  sets. Further, if a critical number $x \in S$ appears in a multiset
  $T \in \mathcal{S}$ of cardinality exceeding 2, then we can split
  $T$ in any way so that $x$ is put in a multiset $T'$ of cardinality
  $2$. It then holds that $\sum T' > v(\bar{S})$, so the resulting
  partition remains an improving partition.

\item There is a bijection $\pi : C \rightarrow L$ such that $\{x, \pi(x)\} \in \mathcal{S}$ (i.e., $C$ is ``matched'' with $L$ in $\mathcal{S}$).
  
  Indeed, by Property 3, every critical number is in a set of
  cardinality 2. Now, let $x$ be a critical number and let $\{x,y\}
  \in \mathcal{S}$ be the multiset of cardinality $2$ that contains
  $x$. If $y$ is not in $L$, then $|C| = |L|$ implies that there is a
  number $y' \in L$ that occurs in a multiset $T$ in $\mathcal{S}$
  that does not contain a critical number.  Because $y' \leq y$, the
  operation of swapping $y'$ and $y$ in $\mathcal{S}$ does not
  decrease the number of multisets that sum to at least $v(\bar{S}) +
  1$. So the partition that results after this swap remains an
  improving partition.
\end{enumerate}

We have $v(\mathcal{S}) > v(\bar{S}) = |C_{+}| + |C|$, so by
Properties 1,2, and 4, there is a multiset $T \in \mathcal{S}$ not
intersecting $C_+$, $C$, and $L$, such that $\sum T > v(\bar{S})$.
Hence $\sum S \backslash (C \cup C_{+} \cup L) \geq \sum T >
v(\bar{S}) = |C| + |C_{+}|$. We conclude that if there is an improving
partition, then $L \cap C = \varnothing$ and $\sum S \backslash (C
\cup C_{+} \cup L) > |C| + |C_{+}|$.

Conversely, if $L \cap C = \varnothing$ and $\sum S \backslash (C \cup
C_{+} \cup L) > |C| + |C_{+}|$, then there is an improving partition.
It consists of

\begin{itemize}
\item the singletons, each containing an element of $C_{+}$,

\item the sets of cardinality $2$, each containing a pair of elements from $C$ and
$L$,

\item the multiset $S \backslash (C \cup C_{+} \cup L)$.
\end{itemize}
\HB
\III

The proof of Theorem~\ref{thm:one} is now immediate. 
It is straightforward to compute $C_+$, $C$ and $L$ in polynomial
time. Using the above lemma we can therefore determine in polynomial time whether an
improving partition exists, and find one in polynomial time if it does.  
\HB 
\III

\NI
\textbf{Proof of Theorem~\ref{thm:two}}. The problem is clearly in $\mathsf{NP}$, so the proof will focus on establishing $\mathsf{NP}$-hardness. We do this by means of a polynomial time reduction from a strongly $\mathsf{NP}$-complete problem. The reduction is from the 3-PARTITION problem. In the 3-PARTITION problem, we are given a multiset $M$ of $3m$ positive integers, such that $\sum M = mb$ for some $b \in \mathbb{N}$. We have to decide whether it is possible to partition this set into $m$ submultisets, such that the sum of the numbers in each submultiset is exactly $b$.

Garey and Johnson \cite{gareyjohnson} prove that the 3-PARTITION problem is strongly $\mathsf{NP}$-complete, even under the assumption that $M$ is represented as above (i.e., non-concisely). This means that the 3-PARTITION problem is $\mathsf{NP}$-complete even when $b$ is bounded by some polynomial in $m$. Denote this polynomial by $p(m)$. From now on, with the SPECIAL 3-PARTITION problem we will mean the special case of the problem where $b$ is bounded by $p(m)$.

Before proceeding, one note is in order. In the original definition of
the 3-PARTITION problem, the additional requirement is imposed that
all sets in the partition are of cardinality 3 (and this is also where
the name of the problem originates from). For convenience, we do not
impose this requirement here. The reason it is not necessary to impose
this requirement is because in \cite{gareyjohnson}, it is shown that
strong $\mathsf{NP}$-hardness holds even when all numbers in the
multiset are strictly between $b/2$ and $b/4$. This enforces that all
sets in the partition will be of cardinality 3. Without the
cardinality constraint, the problem thus becomes more general, and is
automatically strongly $\mathsf{NP}$-hard.

Given a SPECIAL 3-PARTITION instance $(S',m,b)$, we reduce it to an
H-index manipulation problem instance $(S,k)$ as follows. First,
obtain $S''$ from $S'$ by adding $m$ to each number in $S'$. Note that
$(S'',m, k)$, where $k=b+3m$, is a YES-instance of 3-PARTITION if
and only if $(S',m,b)$ is a YES-instance of SPECIAL 3-PARTITION. Note
also that $k-m = b+2m > 0$. Next, obtain the multiset $S$ from $S''$ by
adding $k-m$ copies of $k$ to $S'$. This takes polynomial time, as
$k$ is bounded by $p(m)+3m$. 

We now show that $(S,k)$ is a YES-instance of the H-index manipulation
problem if and only if $(S'',m,k)$ is a YES-instance of 3-PARTITION.

If $(S'',m,k)$ is a YES-instance of 3-PARTITION,
then let $\mathcal{T}$ be a certificate for that, so $\mathcal{T}$ is
a partition of $S''$ into $m$ multisets such that the
sum of the numbers in each multiset is $k$. Then by adding to
$\mathcal{T}$ exactly $k - m$ copies of the set $\{k\}$, we obtain a
certificate that $(S,k)$ is a YES-instance of the H-index
achievability problem, because $k = k$.

Conversely, if $(S,k)$ is a YES-instance of the H-index achievability
problem, then let $\mathcal{T}$ be a certificate for that. We can
assume without loss of generality that the partition $\mathcal{T}$
contains exactly $k - m$ copies of the set $\{k\}$. Indeed,
otherwise we can split each non-singleton set in $\mathcal{T}$ that
contains a copy of $k$ into singleton sets. This will result in
a desired certificate.

By removing all singleton sets $\{k\}$ from $\mathcal{T}$ we obtain a
partition $\mathcal{T}'$ of $S''$. By the choice of $(S,k)$ this new
partition $\mathcal{T}'$ contains $m$ multisets, each of which sums up
to $k$. $\mathcal{T}$ does not contain any additional multiset besides
these $m$ multisets, as then we would have $\sum S'' > mk$, which is
not the case by construction.  Therefore, $\mathcal{T}'$ is a
certificate that $(S'',m,k)$ is a YES-instance of 3-PARTITION.
\HB 
\III

\bibliographystyle{abbrv}
\bibliography{/ufs/apt/bib/e}

\begin{thebibliography}{1}

\bibitem{BK11a}
C.~Bartneck and S.~Kokkelmans.
\newblock Detecting h-index manipulation through self-citation analysis.
\newblock {\em Scientometrics}, 87(1):85--98, 2011.

\bibitem{gareyjohnson}
M.~R. Garey and D.~S. Johnson.
\newblock {\em Computers and Intractability: A Guide to the Theory of
  NP-Completeness}.
\newblock W. H. Freeman, 1979.

\bibitem{Hir05}
J.~E. Hirsch.
\newblock An index to quantify an individual's scientific research output.
\newblock {\em Proceedings of the National Academy of Sciences of the United
  States of America}, 102(46):16569--16572, 2005.

\bibitem{Ste07}
F.~Steutel.
\newblock Persoonlijke ervaringen met de {Hirsch-index}.
\newblock {\em Nieuw Archief voor Wiskunde}, 5:194, 2008.
\newblock In Dutch.

\bibitem{Woe08}
G.~J. Woeginger.
\newblock An axiomatic characterization of the {Hirsch-index}.
\newblock {\em Mathematical Social Sciences}, 56:224--232, 2008.

\bibitem{ZFBE12}
M.~Zuckerman, P.~Faliszewski, Y.~Bachrach, and E.~Elkind.
\newblock Manipulating the quota in weighted voting games.
\newblock {\em Artif. Intell.}, 180-181:1--19, 2012.

\end{thebibliography}
\end{document}